\documentclass[prb,twocolumn,showpacs,floats,eqsecnum,amsmath,amssymb]{revtex4}
\usepackage[dvips]{graphicx}

\begin{document}


\title {Effect of the duration of the synaptic activity on a delayed recurrent neuronal loop}
\author {A. Valizadeh, M. Hashemi, and Y. Azizi}

\affiliation{Institute for
Advanced Studies in Basic Sciences, P.O. Box 45195--1159, Zanjan, Iran}


\begin{abstract}
A recurrent loop consisting of a single neuron is considered which is influenced by a chemical excitatory delayed synaptic feedback. We show the response of the system is dependent to the duration of the activity of the synapse which is determined by the deactivation time constant of the synapse. We show that loops with slow synapses, those which the effect of the synaptic activation remains for time constants comparable to the period of firing, show more predictable results where the effect of the fast synapses is tightly dependent on the loop delay time. The results are compared to those of the loops with inhibitory synapses and also with electrical synapses.
\end{abstract}

\vspace{2mm} \pacs{87.19.lr, 87.19.lg, 87.19.lm}

\maketitle
\section{Introduction}
In control engineering, feedback loops are classical tools to stabilize linear systems\cite{ogata} or suppress chaos in nonlinear systems\cite{femat}. Feedback loops are also abundant in the nervous system, both with the several neurons in the {\it heteroclinic} path or even {\it monoclinic} loops with self-communication via autapses\cite{autapse}; Recurrent networks are known as base structure for creating short and long-term memory\cite{memory} and the delayed feedback loops have been proposed to control or enhance coherent behavior of a group of oscillators\cite{pikovsky,dhamala}.

In neural communication, due to the finite speed of the data transfer in the axons and dendrites, and possible latency in the synapses, communicating between the different area may take delays from few to hundreds of milliseconds, so significant, comparing to the time scales of the neuronal activities. Coherent oscillations of the distance brain area in the brain despite to the notable delay in communication is one of debating problems in the brain context\cite{distantSync1,distantSync2} and due to the importance of the synchronization of the neural population in the biological processes\cite{leon}, cognitive mechanisms\cite{cognition} and for pathological purposes\cite{epilepsy}, the studies devoted of synchronization of the nonlinear oscillators with delayed couplings finds application in neuroscience\cite{strogatz}.

  In this study we consider a model loop containing just a single neuron with a chemical feedback synapse. Auto-synapses in the brain are supposed to be mostly electrical, so most of studies for single neuron with delayed feedback are focused on this type of synapses\cite{massoler,phil,foss}. Yet, a chemical feedback synapse, can be imagined by eliminating other possibly present neurons in the modeling the loop. The main difference is that electrical synapses effect is instantaneous where in chemical synapses activation and deactivation of the synapses have wide range of time constants\cite{gerstner}. Our focus is to study the behavior of the system due to the time constants of the synaptic deactivation and compare the results with those of an electrical synapse. We will show slow and fast synapses, may result in different qualitative effects on the system, when the delay time is varied. We check if the effect of the electrical synapses, because of instantaneous response, is similar to the fast synapses. We also compare the cases of the excitatory and inhibitory feedbacks and discuss the similarities and differences, again due to the synapse time constants.

\begin{figure}[ht!]
\vspace{-.2cm}\centerline{\includegraphics[width=8cm]{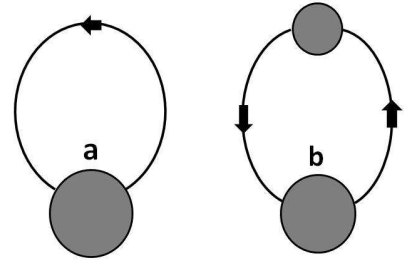}}

\vspace{-.8cm} \caption{ A recurrent loop consisting a single (a) and two neurons (b).} \vspace{0cm} \label{fig1}
\end{figure}

\section{The model}

The main study in this article is based on the classic Hodgkin-Huxley (HH) neuron\cite{HH} with a feedback whose membrane voltage is described by:
\begin{equation}
c\frac{dv}{dt}+I_{na}+I_{k}+I_{l}+I_{syn}=I_{ext},
\end{equation}
$c$ is the capacitance per unit area of the membrane which is taken $1\mu F/cm^{2}$ and $I_{ext}$ holds for the external current. $I_{l}=g_{l}(v-E_{l})$ is the passive leak current and  $I_{na}=g_{na} m^3 h (v-E_{na})$ and  $I_{k}=g_{k} m^4 (v-E_{k})$ are Sodium and Potassium currents respectively. $g_{l}=0.3mS/cm^2$ is the conductance for the leak current and $g_{na}=120mS/cm^2$ and $g_{k}=36mS/cm^2$ are the maximum conductances for the Sodium and Potassium ions, and $E_{l}=10.6mV$, $E_{na}=120mV$ and $E_{k}=-12mV$ are reversal voltages for the leak, Sodium and Potassium currents respectively. $m $ $(h)$, activation (deactivation) variable of Sodium and $n$, activation variable of potassium obey differential equation:
\begin{equation}
\frac{dn_{i}}{dt}=\alpha_{i}(1-n_{i})-\beta_{i}n_{i},
\end{equation}
where $n_{i}$ stands for $m$, $h$ and $n$ respectively and $\alpha$ and $\beta$ are functions of membrane voltage as can be found in Ref. [].

With an electrical synapse, the synaptic current is described by $I_{syn}=g_{el}[v(t)-v(t-\tau)]$ where $g_{el}$ is the synaptic conductivity and $\tau$ is the the loop delay time. With a chemical synapse the feedback synaptic current is described by $I_{syn}=g_{syn} s(t-\tau) (v-E_{syn})$ where $g_{syn}$ is the synaptic maximum conductivity, $E_{syn}$ is the synaptic reversal potential and $\tau$ is the delay of the feedback loop. $s(t)$ is the synaptic activity function defined via:
\begin{equation}
\frac{ds}{dt}=\alpha f(v-v_{th}) (1-s)-\beta s,
\end{equation}
with $\alpha$ and $\beta$ defining the activation and deactivation time constants, $v_{th}=20 mV$ is the threshold voltage for the activation of the synapse and $f$ is a threshold function which is taken $f(x)=1/2(1+\tanh \eta x)$ along this article, where $\eta$ determines how strict is the threshold function. The Parameters are so that the resting potential of the neuron is zero; then $E_{syn}=80 mV$ for excitatory neurons and $E_{syn}=0$ for inhibitory neurons. Activation time of the chemical synapses are usually less than the typical time order of the neuronal data transmission, $1ms$. Instead, deactivation time can take a range from few milliseconds to tens of $ms$. So we fix activation time constant $\alpha$ during the paper and vary $\beta$ to model slow and fast synapses.

We also use the Fitzhugh-Nagumo (FN) model\cite{FN} as a prototype for relaxation oscillators on the same feedback system described by:
\begin{eqnarray}
\nonumber
   \frac{dv}{dt}=v-v^3 /3-w-I_{syn}+I_{ext}, \\
   \frac{dw}{dt}=0.08(v+0.7-.8w),
\end{eqnarray}
with the $v$ and $w$ as the fast (voltage) and slow (recovery) variables respectively. The synaptic current is again described by the Eq. 2.3 but since the spikes in the FN model have different amplitude, the synaptic voltage for the excitatory and inhibitory neurons are $E_{syn}=1.2$ and $E_{syn}=-1.2$ respectively. In what follows when it is not explicitly noted, the voltage and current and conductivity are measured in $mV$ and $\mu A/cm^2$ and $mS/cm^2$ respectively, time in $ms$ and firing rates in $1/ms$.

\begin{figure}[ht!]
\vspace{0cm}\centerline{\includegraphics[width=9cm]{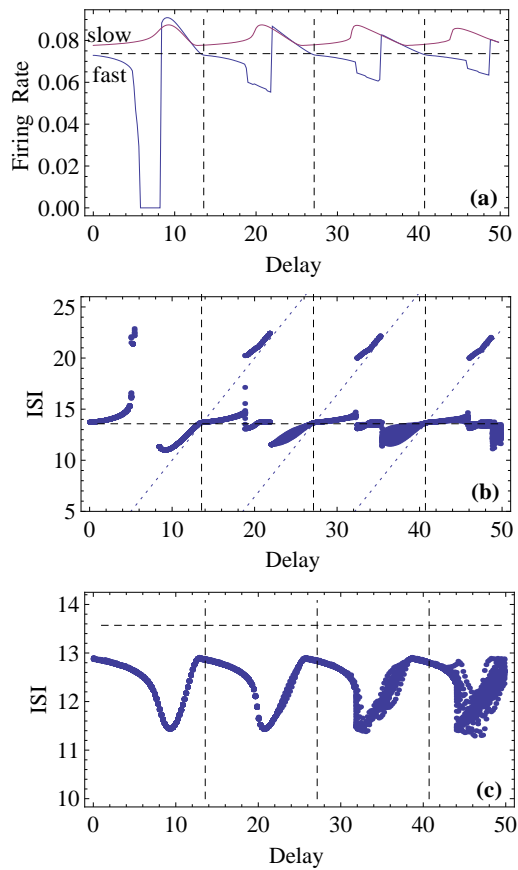}}

\vspace{-1cm} \caption{(color online). (a) Mean firing rate of the HH neuron vs. loop delay time with an excitatory feedback for the fast (blue) and slow (red) synapses. $g_{syn}=0.05$, $\beta=0.5$ for fast and $\beta=0.05$ for slow synapses. The horizontal and vertical dashed lines show the intrinsic rate of firing (without feedback) and the multiples of intrinsic period of firing, respectively. The external current $I_{ext}=6.5 \mu A$ is imposed on the neuron as a step current switched on at $t=0$. (b) and (c) show the inter-spike intervals (bifurcation diagram) with fast and slow synapses respectively. The extra inclined dotted lines in (b) are guide to eye showing when there are ISIs determined by the loop delay time.} \vspace{0cm} \label{fig2}
\end{figure}

\section{Effect of the delayed feedback loop on the firing rate}

In figure 2(a), we have shown the average rate of the firing of the HH neuron vs. the delay time $\tau$ for fast and slow excitatory synapses for the case which the neuron is biased slightly over the threshold for the repetitive firing. It can be seen for the slow synapses, the feedback loop always increases the firing rate of the neuron where for a fast synapse, the effect of the self-synapse is tightly dependent on the delay time. When the loop synaptic pulse arrives in the neuron around the odd multiples of the half delay time, i.e. $\tau \sim (2j+1)T/2$ with $T$ being the period of firing of the open loop neuron ({\it intrinsic period}), the excitatory feedback loop with fast synapse, decreases the firing rate and may even suppress the firing. It also also clear that adding multiples of the intrinsic period of the neuron to the delay, does not result in the similar behavior. For the fast synapses, effect of the larger delays are weaker than the delays below the intrinsic period i.e. variation of the firing rate with respect to open loop firing is smaller for larger delays. For slow synapses, dependence of the firing rate to the delay time is nearly periodic but the period is less than the intrinsic period of the neuron. This is evident as the leftward shift of the maximum of the firing rate for slow synapses, respect to the vertical dashed lines which indicate multiples of the intrinsic period.

\begin{figure}[ht!]
\vspace{0cm}\centerline{\includegraphics[width=10cm]{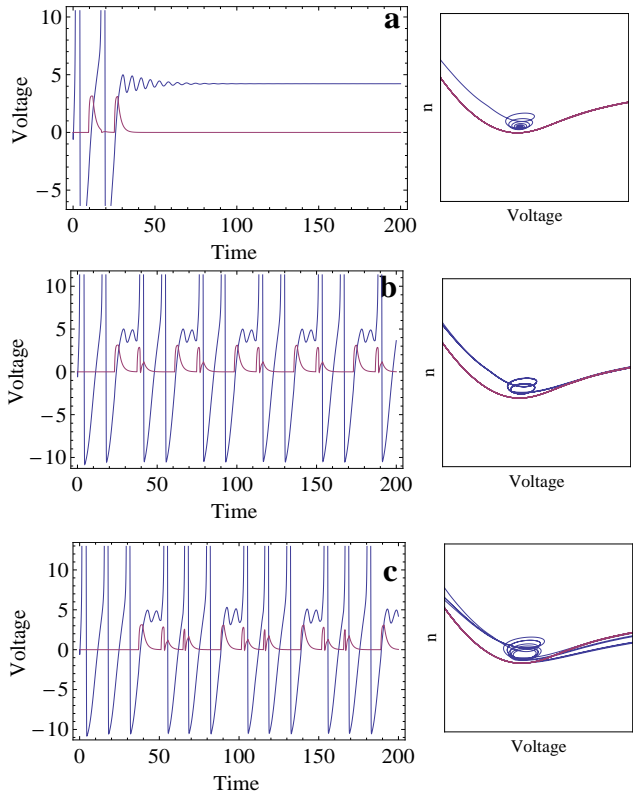}}

\vspace{-3cm} \caption{(color online). In (a) to (c) the evolution of the membrane voltage of the neuron (blue) and he synaptic activity (red) is plotted for three values of delay $\tau=8$, $\tau=T+8$ and $\tau=2T+8$ respectively, where $T=13.57$ is the intrinsic period of the neuron for $I_{ext}=6.5$. The synaptic activity function is exaggerated to be distinguishable. In the right hand side, against each plot for (a) to (c), a reduced phase space representation around fixed points of the system is given.} \vspace{0cm} \label{fig3}
\end{figure}

In Figures 2(b) and 2(c), we have plotted inter spike intervals (ISIs) vs. delay time, a {\it bifurcation diagram}\cite{poincare} for the fast and slow synapses, respectively. With the slow synapse, the only point is the occurrence of aperiodic firing for the loop delay time larger then $2T$\cite{poincare}. With fast synapses for the delay time larger than $T$, three behaviors are recognizable: for delay time near to and greater than multiples of the intrinsic period, the ISIs are not considerably different from intrinsic period i.e. the feedback loop has minor effect on the neuron since all the feedback pulses arrive in the neuron in the refractory period. For delay time near to and less than multiples of the open loop period, the effect of the feedback pulses is a notable decrease of the period of the firing but the interesting behavior is observed when the delay time is about odd multiples of the half delay time $\tau \sim (2j+1)T/2$. In this case ISI may be determined by each of the time constants present in the system: the intrinsic period of the firing of the open loop neuron and the loop delay time. The synaptic activity here consists of groups with two or more pulses, where just one of them leads to an action potential.

\begin{figure}[ht!]
\vspace{-1.5cm}\centerline{\includegraphics[width=11cm]{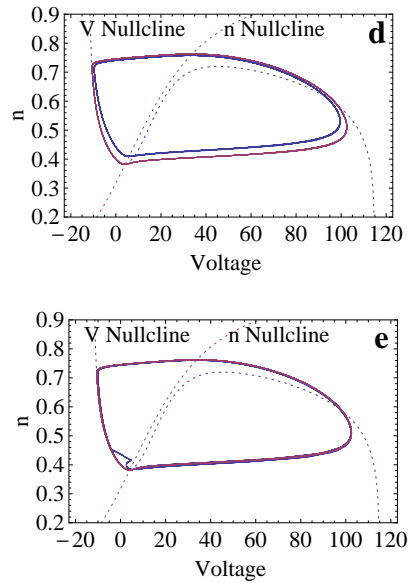}}

\vspace{-5cm} \caption{(color online). In (a) and (b) a the reduced phase space, the membrane voltage vs. potassium gating variable $n$ for the slow (a) and fast (b) synapses is plotted for the full range of the limit cycle. The $v$ and $n$ nullclines for the open loop neuron are plotted with the dotted lines. In the both figures limit cycle of the open loop neuron is also plotted with red lines. For the fast synapse, the deviation from the open loop limit cycle is limited to an excursion seen in the left-bottom of the limit cycle.} \vspace{0cm} \label{fig3}
\end{figure}

Here we try to inspect the origin of the above behaviors, seen in Fig 2. In Fig 3, we have shown the time evolution of the membrane voltage for the system with fast synapse, for three values of delay which differ in an intrinsic period time. A reduced representation of the phase space is also given retaining the voltage $v$ and recovery variable $n$, which are usually retained in the two dimensional reduced versions of HH model\cite{izh}. For the value of the current input we used in the Fig. 3, the HH neuron shows bistability as a result of coexistent of a limit cycle and a stable fixed point. In Fig. 3(a) the pulse suppresses the spiking by sending the phase point to the domain of attraction of the stable fixed foci. For the delays larger than intrinsic period, before the delayed feedback pulse arrives in neuron, the neuron fires an extra action potential, so the delayed feedback consists of a doublet with the interval equal to intrinsic period. When the first pulse of the doublet is going to cease firing as before, the next can shoot the phase point out of the basin of the foci and lead to an action potential as can be seen in Fig. 3(b). Yet, small amplitude oscillations around the foci lags the next spike and reduces the rate of the firing. The action potentials here are in the form of doublets separated by the loop delay time, where the interval between two pulses of doublet is equal to the intrinsic firing period. This scenario is also repeated by the triplets when the delay time is larger than twice of the intrinsic period (Fig. 3(c)).

With a current pulse input, $v$ nullcline would have an upward temporary shift. As a simplified picture, if the phase point of the system can scape to the right hand side of the $v$ nullcline during the its short translocation, the action potential is advanced i.e. an increase in firing rate. This occurs when the phase point is very close to the bottom-left corner of the limit cycle, when the feedback pulse arrives. Otherwise, when the system phase point is on the left arm of the $v$ nullcline, not so close in the bottom, a synaptic pulse may cause the system point to locate in the middle of the left valley of the $v$ nullcline where the evolution is slow due to the existence of both stable and unstable fixed points in this region. Even if the system point is not trapped by the stable fixed point, this temporary excursion out of the limit cycle lags the next action potential resulting in a decrease in the firing rate (see Fig. 4(b)).

 For slow synapses, since the deactivation time constant of the synapse is comparable to or longer than the ISI of the open-loop neuron, the results are different. In this case, after the upward jump of the $v$ nullcline, the return is done slowly. So the system point on the phase space follows the path on the left side of the down moving $v$ nullcline and arrives in the left knee before the nullcline catches the resting location. This always leads to an advance in the occurrence of the action potential since effectively decreases the refractory period by the contraction of the limit cycle height, as it shown in Fig. 4(a). Note that the delay time chosen for Fig. 4(a) and 4(b), is so that the fast and slow synapses, have different effect on the firing rate as can be seen in Fig. 2(a).

\begin{figure}[ht!]
\vspace{-0cm}\centerline{\includegraphics[width=9.5cm]{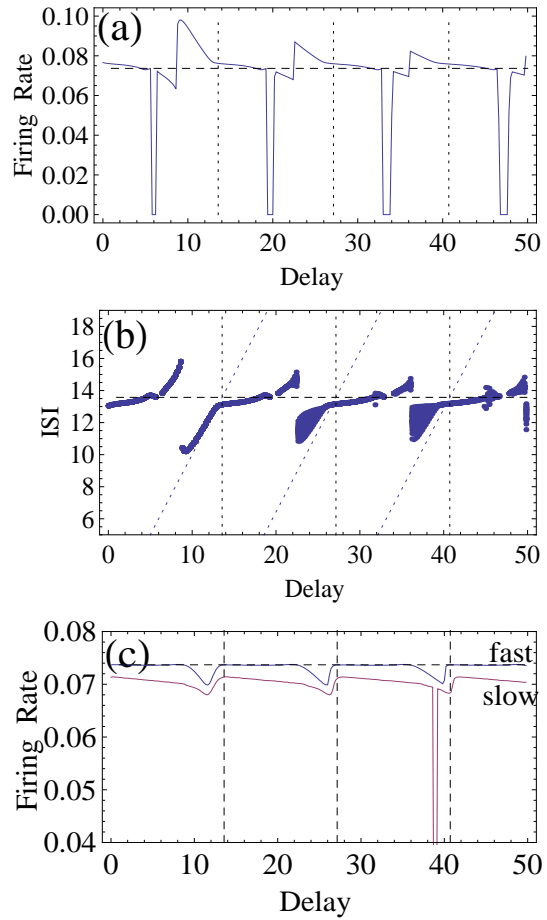}}

\vspace{-0cm} \caption{(color online). (a) Mean firing rate and (b) inter-spike intervals of the HH neuron vs. loop delay time with a feedback via an electric synapse. $g_{el}=0.05$ and the external current $I_{ext}=6.5$ is imposed on the neuron as a step current switched on at $t=0$. In (c) mean firing rate of neuron is plotted when the feedback loop is inhibitory.} \vspace{0cm} \label{fig2}
\end{figure}

For an electrical synapse, since both activation and deactivation are instantaneous, we expect the results to be more similar to those of the fast chemical synapse. As we see in Figs. 5(a) dependence of the firing rate to the loop feedback delay, has similar features with the fast chemical synapse but the bifurcation diagram shown in Fig. 5(b) shows different behavior comparing to Fig. 2(b). Although for the parameters we have chosen the two diagrams are different, our arguments above does not rule out the possibility of the similar behavior for a system with electrical synapses. So if such differences are generic or they can be removed by an suitable choice of parameters, is yet an open question.

We have also checked the system with an inhibitory synapse. As it is seen in Fig. 5(c) the results for both fast and slow chemical synapses are qualitatively similar, i.e. there is no increase of firing rate even when an inhibitory fast synapse is assumed. Suppression of firing here also may occur, but with a slow synapse and for a very narrow range of parameter comparing to the system with a fast excitatory synapse. The bifurcation diagram for such a system (not shown) shows the inhibitory feedback loop with a single neuron never leads to an aperiodic firing. This result worths to note since it seems the loop delay time can not be revealed in the dynamics of the system as we will see also in the next section.

Since the effect of the synaptic feedback on the firing rate is dependent on the geometry of the nullclines and consequently on the behavior of the system on the limit cycle, it is expectable for simplified models which inherit the main properties of the HH phase space, to show similar behavior as the HH model for such study. The simplest model which saves the geometry of the nullclines of the HH equations, is the FH model introduced by Eq. 2.3. We repeated the study of the firing rate change, with FN neurons; it is seen that the behavior of the system under influence of the fast and slow synapses is similar to those of the HH model which supports the arguments about the origin of the phenomenon. We note in passing that for a leaky integrate-fire (LIF) neuron, decrease in the firing rate does not occur for an excitatory fast synaptic feedback. For slow synapses, there is a transient state where the residue of the synaptic activity after each firing, results in an increase in firing rate due to the cumulative effect of synaptic current (results not shown). Even if the LIF model is enriched by an absolute refractory time, just for the values of the delay time where the feedback pulses arrive in the neuron in refractory period, the system dynamic would remain intact but no decrease of the firing rate is seen; confirming the role of the nullclines geometry in the results discussed above.

\begin{figure}[ht!]
\vspace{0cm}\centerline{\includegraphics[width=9.5cm]{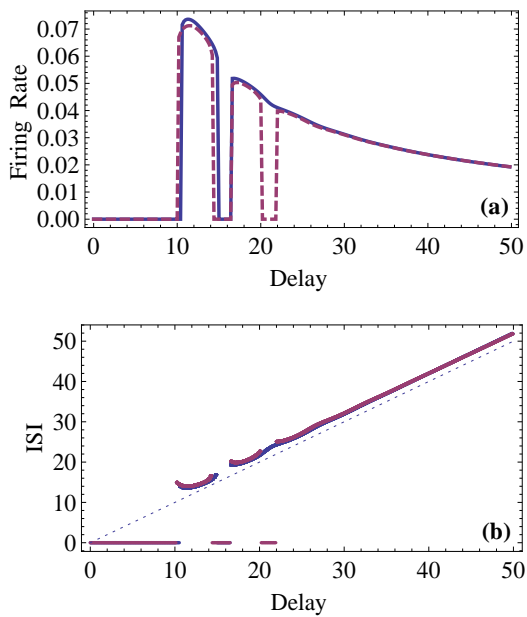}}

\vspace{-4.5cm} \caption{(color online). (a) Average firing rate and (b) inter-spike intervals of the neuron when it is biased by a subthreshold step current $I_{ext}=4.5$ for a fast (thick line) and slow (dashed line). The dotted line in (b) again shows how is the relation of ISI with the loop delay time. Other parameters are those used in Fig. 2.} \vspace{0cm} \label{fig3}
\end{figure}

We end this section with a study of the behavior of system under influence of the feedback when the open-loop neuron is biased below the threshold for repetitive spiking. The results are shown in Fig. 6 with the input current chosen as it causes a single spike at $t=0$. It is shown that the behavior of the system for both fast and slow synapses is in a tight dependence with the delay. It is expectable that neuron remains inactive for small values of delay since the feedback pulse arrives in the refractory period. The threshold delay time for repetitive spiking decreases with increasing the strength of the synapse. For larger value of delay there are again regions with zero activity, arising from the existence of subthreshold oscillation for the HH neuron. For the neurons with sustained subthreshold oscillations, it is known that exact timing of the arrival of the feedback influence determines the behavior of the neuron\cite{massoler,phil,foss}. Here we see for damped subthreshold oscillation, again the feedback timing is detected by the neuron activity. The sensitivity of neuron in the after spike oscillating period varies with the period of such oscillations for both fast and slow synapses. If the neuron is in the minimum of such oscillations, the feedback induced input may not suffice to touch the voltage threshold for the action potential and the neuron is not activated. We note here, for the HH model with subthreshold oscillations, when there is no repetitive firing, the period of these oscillation determines the intrinsic time scale which interacts with the feedback induced time constant and determines the behavior of the system. ISI plot in Fig. 6(b) shows that although period of subthreshold oscillations influences the dynamics of the neuron, the ISI is solely determined by the loop time delay once the neuron is the repetitive spiking state.

\section{delay induced resonant steps}

 If the open loop neuron is in the resting state, due to the initial conditions, a strong enough feedback loop may excite the neuron to fire repeatedly. In this case the feedback delay determines solely the time constant of the activity of the system as it is seen in the Fig. 6. If we fix the delay time constant, this fact will result in plateaus in the firing rate-input current characteristic of the neuron which can be called {\it delay induced resonant steps}. In Fig. 7 we have shown such a characteristic for the system with the electrical synapse and also with fast and slow chemical synapses. It can be seen all types of the synapses may cause delay induced steps on the characteristic on which the firing rate is determined by the delay time $f_{d}=1/ \tau$. Higher order steps steps on the multiple of the $f_{d}$ can also occur but as it is seen they are smoothed and eventually disappeared for the system with slow synapses.

\begin{figure}[ht!]
\vspace{0cm}\centerline{\includegraphics[width=10cm]{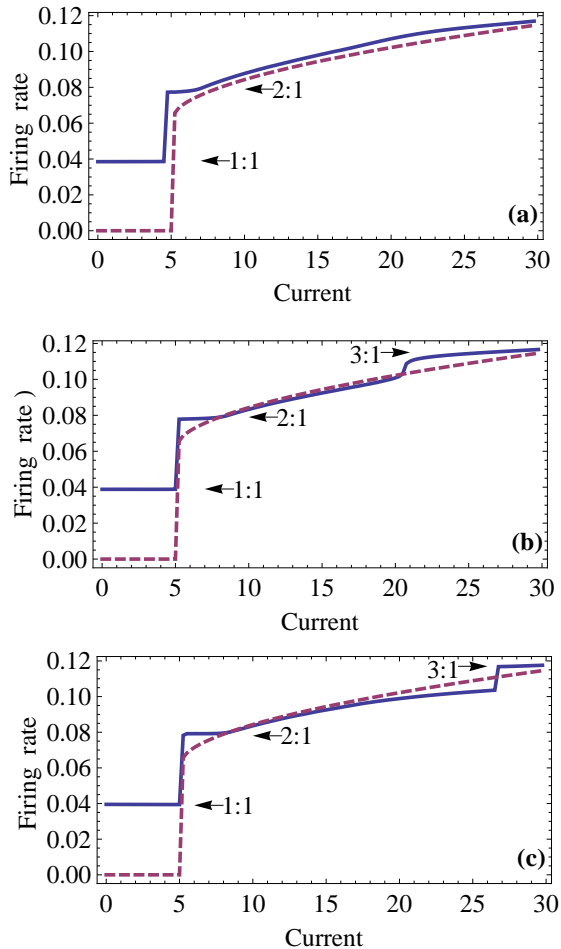}}

\vspace{-0cm} \caption{(color online). Characteristic of the system, average firing rate of the neuron vs. input current is plotted for the open loop neuron (dashed) and in presence of delayed feedback with (a) slow (b) fast and (c) electrical synapses. The time delay is set $\tau=25ms$ and $g_{el}=g_{syn}=0.05$.} \vspace{0cm} \label{fig3}
\end{figure}

 The appearance of the first step depends on the initial condition; existence of a single action potential in the interval $[-\tau,0]$ leads to the first step but the other steps can be seen even if the neuron is in the rest state during the initial period. In general the order of the resonant steps is determined by the number of the effective feedback pulses (those which result in an action potential in every time windows equal to $\tau$: on the second step, before the feedback pulse arrive in neuron the neuron fires another action potential which leads to the doublets for each feedback period. Here despite to what is shown in Fig. 3(b), both the feedback pulses of doublets lead to action potential, resulting in a firing rate equal to $2f_{d}$ and so on for other existing steps. We just mention such steps are not seen in the characteristic of the system with inhibitory synapses, it worths to note again in such system, the loop time constant is not revealed when the feedback is inhibitory.

 These steps, are generic for the all the neuronal models; e.g. they can be seen in the characteristic of a simple LIF neuron. For the simple model studied here, the pulses produced by the neuron circulate in the loop and act as a periodic input on the neuron itself where can entrain the dynamic of the neuron. Considering the effect of the delayed feedback loop as an external periodic force, brings the notion of {\it external synchronization}\cite{sync} in which a nonlinear oscillator is entrained by the external force. Such a phenomena is well-known for the Josephson junctions as the Shapiro steps\cite{shapiro} and also in the loops of coupled Josephson junctions, where the excitations move along the loop as solitary waves, they serve as an extra periodic force on the components of the system and similar resonant steps can be observed\cite{jos}.

\section{conclusion}

Delayed feedback loops enter a new time scale into the dynamics of the system, determined by the delay time. In the neural systems with the chemical synapses, the connections switch on almost instantaneously but they remain active in the time scales which may be comparable to the intrinsic time scales of the neurons and the delay induced time scale. In this work we showed that the duration of synaptic activity is also important in determining the system behavior. Defining fast and slow synapses in reference with the intrinsic time constant(s) of the neuron, it was shown that different qualitative behaviors may be demonstrated by the feedback system. specifically, the effect of the slow synapses for the excitatory synapse, is always speeding up the firing where the effect of the fast synapses is tightly dependent on the delay time. This behavior for the classical HH model, is a consequence of the special characteristic of the limit cycle and the topology of the nullclines in the phase space of the model and until these main characteristic of the HH model is retained in a simplified version, the results will be similar. It is shown behavior of the system with electrical synapses is similar to the systems with fast chemical synapses as is expected. 

With the loops with inhibitory synapses, the observations does not show any strict dependence of the behavior of the system to the activity time of the synapse and the feedback loop via inhibitory synapses always slows down the rate of the firing. Also the system can not be entrained by the inhibitory pulses in the loop despite for the excitatory pulses which can synchronize the firing of the neuron which is revealed as plateaus in the plots of firing rate vs. the constant input current. 

In a more realistic model for the loops containing of more than a single neuron (Fig. 1(b)), our results show when both the bottom-up and top-down synapses are from a single type, say fast or slow, the behavior of the system would be similar to that of a one neuron loop as is expectable. With dissimilar synapses i.e. when one of the synapses is fast and one of them is slow, the dependence of the firing rate to the delay time is similar to the loop with a single neuron with a slow synapse (results not shown). It seems that in a loop consisting of several neurons firing, the faster wins to control the behavior of the system. Here the neuron with slow afferent synapse, would fire faster and would lead the dynamic of the other neuron(s) in the loop. 

\section{Acknowledgement}
The authors acknowledge support by the Institute for Advanced Studies in Basic Sciences (IASBS) Research Council under grant No. G2009IASBS109.

\end{document}